\documentclass[twocolumn,showpacs,aps,prl]{revtex4}

\usepackage{graphicx}
\usepackage{dcolumn}
\usepackage{amsmath}
\usepackage{epsfig}
\long\def\inst#1{\par\nobreak\kern 4pt\nobreak
    {\itshape #1}\par\vskip 10pt plus 3pt minus 3pt}
\RequirePackage{xspace}
\usepackage{relsize}

\def\babar{\mbox{\slshape B\kern-0.1em{\smaller A}\kern-0.1em
    B\kern-0.1em{\smaller A\kern-0.2em R}}}
\def\Kbar    {\kern 0.18em\overline{\kern -0.18em K}{}\xspace}

\def\Kz      {\ensuremath{K^0}\xspace}
\def\Kzb     {\ensuremath{\Kbar^0}\xspace}
\def\KzKzb   {\ensuremath{\Kz {\kern -0.16em \Kzb}}\xspace}

\def\Ks     {\ensuremath{K_S}\xspace}
\def\Kl     {\ensuremath{K_L}\xspace}
\def\KsKs   {\ensuremath{\Ks {\kern -0.16em \Ks}}\xspace}
\def\KlKl   {\ensuremath{\Kl {\kern -0.16em \Kl}}\xspace}
\def\KsKl   {\ensuremath{\Ks {\kern -0.16em \Kl}}\xspace}
\def\KlKs   {\ensuremath{\Kl {\kern -0.16em \Ks}}\xspace}
\def\Dbar    {\kern 0.18em\overline{\kern -0.18em D}{}\xspace}
\def\Dz      {\ensuremath{D^0}\xspace}
\def\Dzb     {\ensuremath{\Dbar^0}\xspace}
\def\DzDzb   {\ensuremath{\Dz {\kern -0.16em \Dzb}}\xspace}
\def\Bbar    {\kern 0.18em\overline{\kern -0.18em B}{}\xspace}

\def\Bz      {\ensuremath{B^0}\xspace}
\def\Bzb     {\ensuremath{\Bbar^0}\xspace}
\def\BzBzb   {\ensuremath{\Bz {\kern -0.16em \Bzb}}\xspace}
\def\Bu      {\ensuremath{B^+}\xspace}
\def\Bub     {\ensuremath{B^-}\xspace}

\def\BpBm    {\ensuremath{\Bu {\kern -0.16em \Bub}}\xspace}

\newcommand{\optbar}[1]{\shortstack{{\tiny (\rule[.4ex]{1em}{.1mm})}
  \\ [-.7ex] $#1$}}
\def\BorBbar    {\kern 0.18em\optbar{\kern -0.18em B}{}\xspace}
\def\DorDbar    {\kern 0.18em\optbar{\kern -0.18em D}{}\xspace}
\def\KorKbar    {\kern 0.18em\optbar{\kern -0.18em K}{}\xspace}

\def\pep2{PEP-II}
\mathchardef\Upsilon="7107
\def\Y#1S{\ensuremath{\Upsilon{(#1S)}}\xspace}

\begin{document}

\title{
\large \bfseries \boldmath Study of the electromagnetic transitions
$J/\psi \rightarrow P l^+l^-$ and probe dark photon}

\author{Jinlin Fu$^{1,2}$}\email{fujl@stu.nju.edu.cn>}
\author{Hai-Bo Li$^2$}\email{lihb@mail.ihep.ac.cn}
\author{Xiaoshuai Qin$^2$}\email{qinxs@mail.ihep.ac.cn}
\author{Mao-Zhi Yang$^3$}\email{yangmz@nankai.edu.cn}
\affiliation{$^1$ Department of Physics, Nanjing University, Nanjing 210093, China \\
 $^2$ Institute of High Energy Physics, P.O.Box 918, Beijing  100049, China \\
 $^3$ School of Physics, Nankai University, Tianjin, 300071, China}


\date{\today}


\begin{abstract}
We study the electromagnetic Dalitz decay modes of $J/\psi
\rightarrow P l^+l^-$ ($P= \pi^0$, $\eta$ or $\eta^\prime$). In
these decays, the lepton pairs are formed by internal conversion of
an intermediate virtual photon with invariant mass $m_{l^+l^-}$.
Study of the effective-mass spectrum of the $l^+l^-$ will shed light
on the dynamic transition form factor $F_{J/\psi P}(q^2)$ ($q^2 =
m^2_{l^+l^-}$), which characterizes the electromagnetic structure
arising at the vertex of the transition $J/\psi$ to pseudoscalars.
We also discuss the direct productions of a GeV scale vector $U$
boson in these processes $J/\psi \rightarrow PU$ ($U\rightarrow
l^+l^-$). It is responsible for mediating a new $U(1)_d$
interaction, as recently exploited in the context of weakly
interacting massive particle dark matter.
 In this paper, we firstly use the usual pole approximation for
the form factor to estimate the decay rate of $J/\psi \rightarrow P
l^+l^-$ in the standard model.  Then the reach of searching for the
dark photon is estimated. We suggest that these  Dalitz decays can
be used to search for the light $U$ boson in the BESIII experiment
with a huge $J/\psi$ data set.

\end{abstract}


\maketitle

\section{Introduction}

Study of the electromagnetic (EM) decays of the hadronic states is
of interest for understanding the structure of hadronic matter and
for revealing the fundamental mechanisms for the interactions of
photons and hadrons~\cite{landsberg1982,landsberg1985}. The EM
Dalitz decays $V \rightarrow P l^+l^-$ of light unflavored vector
mesons ($\rho$, $\omega$ or $\phi$) are specially interesting for
probing the electromagnetic structure arising at the vertex of the
transition $V$ to the pseudoscalar. In table~\ref{lightdecays}, we
summarize the experimental results for the EM Dalitz decays of the
light vector mesons. The ratios of the Dalitz decays to the
corresponding radiative decays of vector mesons are suppressed by
two orders of magnitude, especially, for $V\rightarrow P e^+e^-$
modes. Assuming point-like particles, the decay rate of this process
vs. $m_{l^+l^-}$ can be exactly described by QED~\cite{kroll1955} in
the standard model (SM). However, the rate is strongly modified by
the dynamic transition form factor $F_{V P}(q^2)$ ($q^2 =
m^2_{l^+l^-}$), which can be estimated based on models of
QCD~\cite{sov1992,faessler2000,klingl,kopp}.

\begin{table}[htbp]
\caption{The experimental results on the light vector Dalitz decays
, $V\rightarrow P l^+l^-$ ($V= \rho$, $\omega$ or $\phi$), and
ratios of the Dalitz decays to radiative decays of the vector
mesons. These data are from
PDG2010~\cite{pdg2010}.}\label{lightdecays}
\begin{tabular}{c|c|c}\hline
 Decay mode  & Experimental results & $\frac{\Gamma(V\rightarrow P l^+l^-)}{\Gamma(V\rightarrow P\gamma)}$  \\ \hline
$\rho^0 \rightarrow \pi^0 e^+ e^-$& $<1.2 \times 10^{-5} $ (90\%
C.L.) & $< 2.0\times 10^{-2}$ \\ \hline $\omega \rightarrow \pi^0
e^+
e^- $ & $(7.7\pm 0.6) \times 10^{-4}$ & $(0.93 \pm 0.08) \times 10^{-2}$ \\
\hline $\omega \rightarrow \pi^0 \mu^+ \mu^-$& $(1.3\pm 0.4) \times
10^{-4}$ & $(0.16\pm 0.05) \times 10^{-2}$\\\hline $\omega
\rightarrow \eta e^+e^-$ & $<1.1\times 10^{-5}$ (90\% C.L.) &
$<2.4\times
10^{-2}$ \\
 \hline $\phi
\rightarrow \eta e^+ e^-$& $(1.15\pm 0.10) \times 10^{-4}$& $(0.88
\pm 0.08)\times 10^{-2}$ \\ \hline $\phi \rightarrow
\eta \mu^+ \mu^-$& $<9.4 \times 10^{-6}$(90\% C.L.) & $<0.07 \times 10^{-2}$\\
\hline
\end{tabular}
\end{table}

Experimentally, $|F_{VP}(q^2)|^2$ is directly accessible by
comparing the measured invariant mass spectrum of the lepton pairs
from the Dalitz decays with the point-like QED prediction. A
comprehensive review of the topic is contained in
reference~\cite{landsberg1985}. Recently, high quality data from
NA60 experiment measured the $q^2$-dependent form factor of the
Dalitz decay $\omega \rightarrow \pi^0 \mu^+\mu^-$~\cite{na60-2009}.
Using the usual pole approximation $F(q^2) =
1/(1-\frac{q^2}{\Lambda^2})$ for the form factor, the $\Lambda^{-2}$
has been found to be $2.24 \pm 0.06$ GeV$^{-2}$, which strongly
deviates from the expectation of vector meson
dominance(VMD)~\cite{iva2011,le2010,faessler2000,klingl,kopp}. The
form factor showed a relative increase close to the kinematic
cut-off by a factor of $\sim$ 10~\cite{na60-2009}. For the decay of
$\phi \rightarrow \eta e^+e^-$, the SND experiment has looked for
the $m_{e^+e^-}$ invariant mass distribution with 213
events~\cite{snd-phi}, and measured the form factor slope
$\Lambda^{-2}$ to be $3.8\pm 1.8$ GeV$^{-2}$. Most recently, KLOE-2
has selected 7000 $\phi \rightarrow \eta e^+e^-$ events with $\eta
\rightarrow \pi^+\pi^-\pi^0$ using a sample of 739 pb$^{-1}$ on the
$\phi$ peak~\cite{kloe-hadron11}. A preliminary fit to the
$m_{e^+e^-}$ indicates the possibility to reach a 5\% error on the
form factor slope~\cite{kloe-hadron11}.

These theoretical and experimental investigations of the EM Dalitz
decays of light vector mesons motivate us to study the rare
charmonium decays $J/\psi \rightarrow P l^+l^-$ ($P= \pi^0$, $\eta$
or $\eta^\prime$). The measurements of the $q^2$-dependent form
factors will provide useful information on the interaction of the
charmonium states with electromagnetic field. In particular, with
more phase space, the transition between $J/\psi$ and the
pseudoscalar can be explored over a large region of momentum
transfer, which will be used to test QCD prediction cleanly. The
decay rates of the $J/\psi \rightarrow \gamma \pi^0$, $\gamma \eta$
and $\gamma \eta^\prime$ are $(3.49^{+0.33}_{-0.30})\times 10^{-5}$,
$(1.104\pm0.034)\times 10^{-3}$ and $(5.28\pm0.15)\times
10^{-3}$~\cite{pdg2010}, respectively. From a direct estimation
according to the ratios of $\omega$ and $\phi$ Dalitz decays to the
corresponding radiative decays, the expected Dalitz decay rates
could reach $\sim 10^{-7}$ for $J/\psi \rightarrow \pi^0 e^+e^-$ and
$\sim 10^{-5}$ for $J/\psi \rightarrow \eta e^+e^-$ and $\eta^\prime
e^+e^-$, respectively.

It is also  interesting to search for dark photon, the light $U$
boson, in $J/\psi \rightarrow P U$ and $U\rightarrow l^+l^-$, in
which the virtual photon is replaced by the light on-shell dark
photon $U$. The light U boson may couple to the SM charged particles
with a much suppressed coupling which has been considered in various
contexts~\cite{dark-1,dark-2,dark-3, dark-4,dark-5,dark-6}. We
consider the new Abelian gauge group $U(1)_d$ which has a
gauge-invariant kinetic mixing with the SM hypercharge
$U(1)_Y$~\cite{u-1,u-2,u-3}. After electroweak symmetry breaking, we
have the Lagrangian:
\begin{equation}
{\cal L} = {\cal L}_{SM} + \epsilon_Y F^{Y,\mu\nu}F^d_{\mu\nu} +
m^2_{U} A^{d, \mu}A^d_{\mu},
 \label{mixingl}
\end{equation}
where ${\cal L}_{SM}$ is the SM Lagrangian, $F^Y_{\mu\nu}$ and
$F^d_{\mu\nu}$ are the field strength for  the SM gauge boson $B$
and $U$ boson, respectively,  $A^d$ is the gauge field of a massive
dark $U(1)_d$ gauge group~\cite{u-2}. The second term in
Eq.(\ref{mixingl}) is kinetic mixing operator, and $\epsilon \sim
10^{-8} - 10^{-2}$ is generated at any scale by loops of heavy
fields charged under both $U(1)$s. In a supersymmetry theory, the
kinetic mixing operator induces a mixing between the $D$-terms
associated with $U(1)_d$ and $U(1)_Y$. The hypercharge $D$-term gets
a vacuum expectation value from the electroweak symmetry breaking
and induces a weak-scale effective Fayet-Iliopoulos term for
$U(1)_d$. Consequently, the $U(1)_d$ symmetry breaking scale is
suppressed by loop factors or by $\sqrt{\epsilon}$, leading to MeV
to GeV-scale $U$ boson mass~\cite{u-1,u-3}. The parameters of
concern in this paper are $\epsilon$ and $m_{U}$.

In the BESIII experiment, more than 1 billion $J/\psi$
($\psi^\prime$) sample will be collected next year, the first look
of these decay modes will be accessible~\cite{haibo-hadron11}. It
will shed light on probing new physics beyond the standard model,
such as possible $U$ boson with mass less than 300 MeV
 range~\cite{fayet1,fayet2,fayet3,li2010,kloe}, which may contribute to
the process by replacing virtual photon. By looking at the spectrum
of the dilepton, with huge data set in the BESIII experiment, one
may see possible new physics contribution which will modify the
shape of the di-lepton spectrum and total decay rate of the Dalitz
decay process.

In this paper, the full angular distribution and $q^2$-dependent
rate are derived in the SM framework in Section~\ref{dalitz}.  Using
the pole approximation, the decay rates for the EM Dalitz decays of
$J/\psi$ are estimated for the first time. An interesting
consequence is that the dark photon can be probed in the low energy
BESIII experiment. In Section~\ref{darkphoton}, we estimate the
reach of the $U$ boson search in the $J/\psi \rightarrow P U$ and
$U\rightarrow l^+l^-$ decay.

\section{$q^2$-dependent decay rate of $J/\psi \rightarrow P l^+l^-$}
\label{dalitz}

 The amplitude of the Dalitz decay $\psi \rightarrow P
l^+l^-$ (hereafter, $\psi$ denotes $J/\psi$) has the
Lorentz-invariant form:
\begin{eqnarray}
T = 4\pi \alpha f_{\psi P}\epsilon^{\mu\nu\rho\sigma} p_{\mu}
q_{\nu} \epsilon_{\rho} \frac{1}{q^2} \bar{u}_1 \gamma_{\sigma}v_2,
 \label{eq:invariance}
\end{eqnarray}
where $f_{\psi P}$ is the form factor of the $\psi \rightarrow P$
transition; $q_\nu$ is the 4-momentum of the virtual photon or the
total 4-momentum of $l^+l^-$ ($l= e$, $\mu$) system ; $q^2 =
m^2_{l^+l^-}$ is the effective mass squared of the lepton pair;
$p_\mu$ is the 4-momentum of the pseudoscalar meson; $\epsilon_\rho$
is the polarization 4-vector of the $\psi$;
$\epsilon^{\mu\nu\rho\sigma}$ is the totally antisymmetric unity
tensor. It is straightforward to obtain the magnitude of the amplitude squared as below:
\begin{widetext}
\begin{eqnarray}
 |T|^2 = 16\pi^2 \alpha^2 \frac{|f_{\psi P}(q^2)|^2}{q^4} \left[
 8(p\cdot q)^2 m_l^2 - 8p^2q^2m^2_l -
  2p^2q^4 -8 (k_1 \cdot p)(k_2\cdot p)q^2+4(p\cdot q)^2q^2\right],
 \label{eq:ampl}
\end{eqnarray}
where $m_l$ is the lepton mass;  $q=k_1 + k_2$, and $p$, $k_1$ and
$k_2$ are 4-momenta of particles $P$, $l^+$ and $l^-$, respectively.

The angular distribution of the differential decay width can be obtained as
\begin{eqnarray}
 \frac{d\Gamma(\psi \rightarrow P l^+l^-)}{d q^2} &=&\frac{1}{3}
 \frac{\alpha^2}{256\pi^3 m_{\psi}^3} \frac{|f_{\psi
 P}(q^2)|^2}{q^2}\left(1-\frac{4m^2_l}{q^2}\right)^{1/2}
 \left[(m^2_{\psi}-m^2_P+q^2)^2-4m^2_{\psi}q^2\right]^{3/2}\times \nonumber \\
 && \int d\Omega_3 d\Omega^*_1 \left[\left(1+\frac{4m^2_l}{q^2}\right)
 + \left(1-\frac{4m^2_l}{q^2} \right ) \mbox{cos}^2\theta^*_1 \right],
 \label{eq:dgammaan}
\end{eqnarray}
where $m_{\psi}$ and $m_P$ are the masses of the initial charmonium
state and pseudoscalar meson; $d\Omega_3 =
d\phi_3d(\mbox{cos}\theta_3)$ is the solid angle of $P$ in the rest
frame of $\psi$ and $d\Omega^*_1 = d\phi^*_1
d(\mbox{cos}\theta^*_1)$ is the solid angle of one of the lepton pair in the rest
frame of $l^+l^-$ system (the $z$ direction is defined as the
momentum direction of $l^+l^-$ in the $\psi$ system); $\theta^*_1$
is the helicity angle of $l^+l^-$ system, which is defined as the
angle between momentum direction of one of the lepton pair and
direction of the $P$ meson in the rest frame of $l^+l^-$ system. By
integrating the solid angles in Eq.~\ref{eq:dgammaan}, one can
obtain the $q^2$-dependent differential decay width:
\begin{eqnarray}
 \frac{d\Gamma(\psi \rightarrow P l^+l^-)}{d q^2} =\frac{1}{3}
 \frac{\alpha^2}{24\pi m_{\psi}^3} \frac{|f_{\psi
 P}(q^2)|^2}{q^2}\left(1-\frac{4m^2_l}{q^2}\right)^{1/2}
\left(1+\frac{2m^2_l}{q^2}\right)
 \left[(m^2_{\psi}-m^2_P+q^2)^2-4m^2_{\psi}q^2\right]^{3/2}.
 \label{eq:dgamma}
\end{eqnarray}
\end{widetext}

For the corresponding radiative decay of $\psi \rightarrow P
\gamma$, the decay width can be obtained as :
\begin{eqnarray}
\Gamma(\psi \rightarrow P \gamma ) = \frac{1}{3}
\frac{\alpha(m^2_{\psi} - m^2_P)^3} {8 m^3_{\psi}}|f_{\psi P}(0)|^2.
 \label{eq:gammap}
\end{eqnarray}
From Eqs.~(\ref{eq:dgamma}) and~(\ref{eq:gammap}) the
$q^2$-dependent differential decay width in the $\psi \rightarrow P
l^+l^-$ decay normalized to the width of the corresponding radiative
$\psi \rightarrow P \gamma$ is derived :
\begin{widetext}
\begin{eqnarray}
 \frac{d\Gamma(\psi \rightarrow P l^+l^-)}{d q^2 \Gamma (\psi \rightarrow P \gamma)}
 &=&
 \frac{\alpha}{3\pi} \left|\frac{f_{\psi
 P}(q^2)}{f_{\psi P}(0)}\right|^2 \frac{1}{q^2} \left(1-\frac{4m^2_l}{q^2}\right)^{1/2}
\left(1+\frac{2m^2_l}{q^2}\right)
 \left[\left(1+\frac{q^2}{m^2_{\psi}-m^2_P}\right)^2 - \frac{4m^2_{\psi}q^2}{(m^2_{\psi}-m^2_P)^2}\right]^{3/2}\nonumber \\
 &=& |F_{\psi P}(q^2)|^2 \times [\mbox{QED}(q^2)],
 \label{eq:dgamman}
\end{eqnarray}
\end{widetext}
where the normalized form factor for the $\psi \rightarrow P$
transition is defined as $
F_{\psi P}(q^2)\equiv f_{\psi P}(q^2)/f_{\psi P}(0) $, and the normalization is $F_{\psi P}(0) =1$.
The form factor defines the electromagnetic properties of the region
in which $\psi$ is converted into pseudoscalar. By comparing the
measured spectrum of the lepton pairs in the Dalitz decay with QED
calculations for point-like particles, it is possible to determine
experimentally the transition form factor in the time-like region of
the momentum transfer~\cite{landsberg1982,landsberg1985}. Namely, the
form factor can modify the lepton spectrum as compared with that
obtained for point-like particles. 

For the decays accompanied by the production of the
electron-positron pair, we should note that the radiative
corrections proportional to $\alpha \mbox{ln}^2(q^2/m^2_l)$ will be
important. We will not discuss the high order QED corrections in
this analysis since the data sample in the BESIII experiment is
still small, and BESIII is expected to see the first signal for the
effect at leading order. In addition to that, the external
conversion of the $\gamma$ from the radiative decay of $\psi
\rightarrow P \gamma$ will make the analysis more complicated,
however, at the BESIII the external conversion rate could be up to
2\%, and the invariant mass of the $m_{e^+e^-}$ will form a narrow
peak at 20-40 MeV, which will not really affect the slope shape of
the dilepton. For the decays accompanied by the production of the
muon pairs the radiative corrections and external radiation effects
are negligibly small.

\begin{center}
\begin{figure}[ht]
\scalebox{0.6}{\epsfig{file=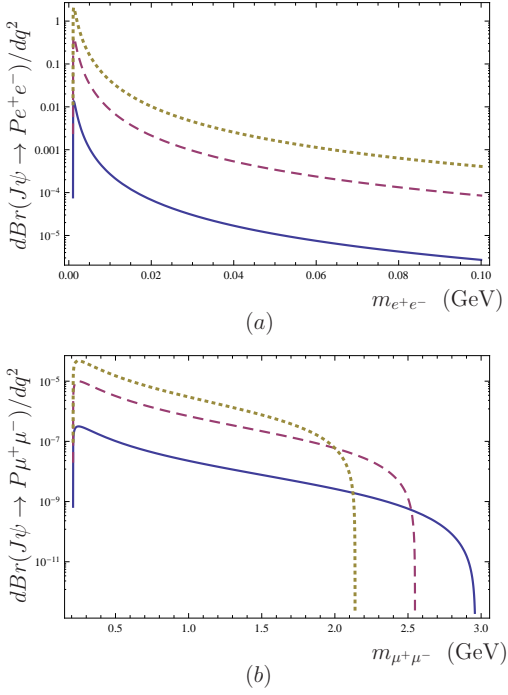}} \caption{The differential
decay rates for $\psi \rightarrow P l^+ l^-$, where the solid curve
is for $\psi \rightarrow \pi^0 l^+ l^-$, the dashed curve for $\psi
\rightarrow \eta l^+ l^-$ and the dotted for $\psi \rightarrow
\eta^\prime l^+ l^-$. (a) is for the case that the lepton pair is
$e^+e^-$, while (b) for the case $\mu^+\mu^-$ . In (a), only part of
the phase space for $m_{e^+e^-}$ is shown for demonstration purpose
since most of the phase space is accumulated near small $m_{e^+e^-}$
region. } \label{fig:dd}
\end{figure}
\end{center}

\begin{center}
\begin{figure}[ht]
\scalebox{0.6}{\epsfig{file=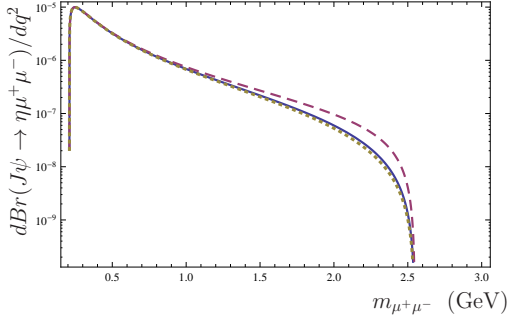}} \caption{The differential
decay rate for $\psi \rightarrow \eta \mu^+ \mu^-$, where the solid curve
is for the pole mass taken as $\Lambda=m_{\psi\prime}$, the dashed curve for
$\Lambda=3.0\;{\rm GeV}$ and the dotted for $\Lambda=4.0\;{\rm GeV}$.} \label{poledepen1}
\end{figure}
\end{center}

\begin{center}
\begin{figure}[ht]
\scalebox{0.6}{\epsfig{file=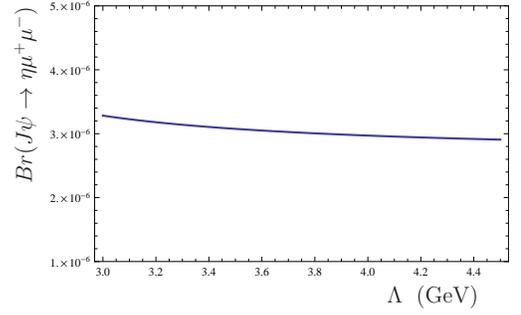}} \caption{The decay rate for
$\psi \rightarrow \eta \mu^+ \mu^-$ with the variation of the pole mass $\Lambda$.} \label{poledepen2}
\end{figure}
\end{center}

 To estimate the order of magnitude, one may use the Vector
 Dominance Model (VDM), in which the hadronic EM current is
 proportional to vector meson fields~\cite{gellman61,bauer78}. Hence
 the VDM predicts a growth of the transition form factors with
 increasing dilepton mass. The form factor may be parameterized in
 the simple pole approximation as
\begin{eqnarray}
F_{\psi P}(q^2) = \frac{1}{1-\frac{q^2}{\Lambda^2}},
 \label{eq:ff}
\end{eqnarray} 
where the pole mass $\Lambda$ should be the mass of the vector
resonance near the energy scale of the decaying particle according
to the VDM model. In $\psi$ decay the pole mass could be the mass of
$\psi^\prime$. By assuming the pole approximation and taking
$\Lambda = m_{\psi^\prime}$, in Fig.~\ref{fig:dd} we show the
differential decay rates for $\psi \rightarrow \pi^0 l^+ l^-$, $\eta
l^+l^-$ and $\eta^\prime l^+l^-$, respectively. The decay rates for
$\psi \rightarrow \pi^0 l^+ l^-$, $\eta l^+l^-$ and $\eta^\prime
l^+l^-$ are estimated and presented in table \ref{t2}. To study the
dependence of the decay rates on the value of the pole mass, we
varied the pole mass. The results for $J/\psi \to\eta\mu^+\mu^-$ are
shown in Figs.~\ref{poledepen1} and \ref{poledepen2} as an example.
The cases for the other decay modes considered in this work are
similar. Both the differential and total decay rates are not
sensitive to the value of the pole mass if the pole mass is in the
range $\Lambda^2>q^2_{\rm max}=(m_\psi-m_p)^2$ in the Dalitz decay
process of $J/\psi$. The reason can be well understood. The dominant
contribution to the decay rate comes from the region of small value
of $q^2$. For the pole mass with large value, $q^2/\Lambda ^2$ is
small, therefore this term cannot give large effect. This is
different from the case for the light vector meson Dalitz decays.

Because the decay rates are not sensitive to the pole mass in the
form factor, the predicted decay rates in Table \ref{t2} are more
reliable. Comparing these predictions with experimental measurement
should make sense.  In the BESIII experiment, more than 1 billion
$J/\psi$ ($\psi^\prime$) sample will be collected in year 2012, the
first look of these decay modes will be
accessible~\cite{haibo-hadron11}.

\begin{center}
\begin{table}[h]
\caption{ The estimated decay rates for $\psi \rightarrow \pi^0 l^+
l^-$, $\eta l^+l^-$ and $\eta^\prime l^+l^-$ based on
Eq.(~\ref{eq:dgamman}) by assuming  pole approximation and $\Lambda
= m_{\psi^\prime}$. The error on the decay rate is from the measured
error of $\psi \rightarrow P \gamma$ which is used as normalization.
}
 \label{t2}
\begin{tabular}{c|c|c}\hline
  Decay mode  & $e^+e^-$ & $\mu^+\mu^-$ \\ \hline
$\psi \rightarrow \pi^0 l^+ l^-$ & $(3.89^{+0.37}_{-0.33}) \times 10^{-7}$ & $(1.01^{+0.10}_{-0.09} )\times 10^{-7}$
\\  \hline
$\psi \rightarrow \eta l^+ l^-$ & $(1.21\pm 0.04 )\times 10^{-5}$ & $(0.30\pm 0.01) \times 10^{-5}$
\\  \hline
$\psi \rightarrow \eta^\prime l^+ l^-$ & $(5.66\pm 0.16)\times 10^{-5}$ & $(1.31\pm 0.04) \times 10^{-5}$ \\
\hline
\end{tabular}
\end{table}
\end{center}

\section{Reach of U-boson search in $J/\psi \rightarrow P U$,
$U\rightarrow l^+l^-$ decay.} \label{darkphoton}

In this section, we discuss the constraints and discovery potential
for the $U$ boson at Beijing electron-positron collider II (BEPCII)
~\cite{bepcii}. Since the $U$ boson couples mainly to the SM
electromagnetic current~\cite{wang}. Its production at BEPCII is the
same as that of photon, although with a much suppressed rate.
Therefore, the process $J/\psi \rightarrow P \gamma^* \rightarrow
Pl^+l^-$ will have a chance to look for U boson as well.

The on-shell U boson will decay to a pair of leptons in $J/\psi
\rightarrow PU$, leading to a signal of $Pl^+l^-$. The SM background
$J/\psi \rightarrow P \gamma^* \rightarrow Pl^+l^-$, although large
for this process, is not a severe problem as the kinematics of the
signal are quite distinct. The invariant mass of the lepton pair is
just within a single bin due to the tiny width of the vector $U$
boson and can be distinguished from the SM background. It will be
interesting to look for a low mass,  up to GeV scale, U boson in
these modes.

\begin{figure}[htb]
\includegraphics[width=0.8\columnwidth]{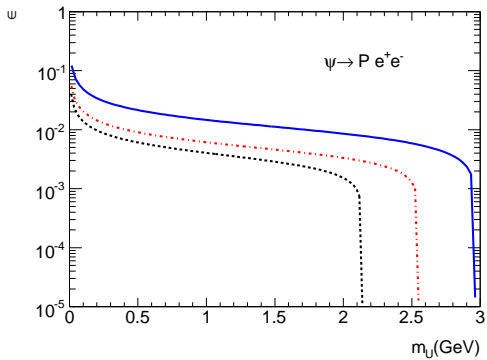}
 \caption{Illustrative plot of the reach of vector boson at BES-III in the channel of
 $\psi \rightarrow PU$ (solid curve for $P=\pi^0$; dot-dashed curve for $P=\eta$ and dashed curve
 for $P=\eta^\prime$, respectively), where $U$ decay
into $e^+e^-$. } \label{eep}
\end{figure}

\begin{figure}[htb]
\includegraphics[width=0.8\columnwidth]{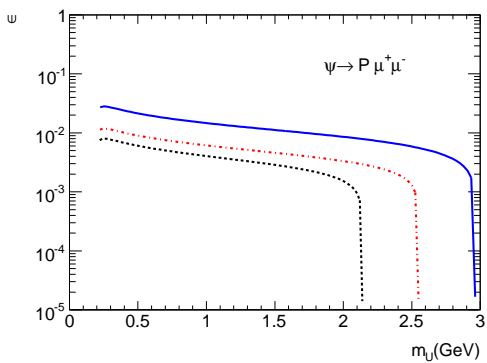}
 \caption{The same as Fig.~\ref{eep} with $U$ decay
into $\mu^+ \mu^-$.} \label{mumup}
\end{figure}

From Eq.~(\ref{eq:dgamman}), the number of background events in the
window of $\delta q^2$ (resolution of the $q^2$) around $q^2 =
m^2_{l^+l^-} = m^2_{U}$ is about
\begin{eqnarray}
 N_B
 &=& N_{\psi}\int^{q^2_{\text{max},i}}_{q^2_{\text{min}, i}} \frac{d\Gamma(\psi \rightarrow P l^+l^-)}{d q^2 \Gamma (\psi \rightarrow P
  \gamma)}dq^2 \times BR(\psi \rightarrow P\gamma)
  \nonumber \\
 &\approx&
  N_{\psi}\frac{d\Gamma(\psi \rightarrow P l^+l^-)}{d q^2 \Gamma (\psi \rightarrow P
  \gamma)}\delta q^2 \times BR(\psi \rightarrow P\gamma)
\nonumber \\
 &=&
 N_{\psi}|F_{\psi P}(q^2)|^2 \times [\mbox{QED}(q^2)]  \delta q^2 \times BR(\psi \rightarrow P\gamma),\nonumber \\
 \label{eq:bgk}
\end{eqnarray}
where $q^2_{\text{max},i}$ ($q^2_{\text{min}, i}$) is the upper
(lower) value of the $i$-$th$ $q^2$ bin; $N_{\psi}$ is the total
number of $\psi$ decay events; and $BR(\psi \rightarrow P\gamma)$ is
the branching fraction from PDG~\cite{pdg2010}. The size of the bin
is the window size $\delta q^2$ which is obtained from the
resolution functions of the $m_{l^+l^-}$ in Eqs. (4) and (5) in
Ref~\cite{li2010} based on the BESIII Monte Carlo simulations. By
replacing the photon by $U$ boson in $\psi\rightarrow P\gamma$, the
signal rate can be estimated to be $BR(\psi \rightarrow PU) \approx
\epsilon^2 BR(\psi \rightarrow P\gamma)$, where $\epsilon$ was
defined in Eq.~(\ref{mixingl}). Thus, the expected number of signal
events is about
\begin{eqnarray}
 N_S = N_{\psi}\times \epsilon^2 BR(\psi \rightarrow
 P\gamma)BR(U\rightarrow l^+l^-),
 \label{eq:sig}
\end{eqnarray}
where we assume $BR(U\rightarrow l^+l^-)=1$ in this study. Combining
Eqs.~(\ref{eq:bgk}) and (\ref{eq:sig}), We estimate the expected
numerical results based on the significance
\begin{eqnarray}
\frac{S}{\sqrt{B}} &=& \frac{N_S}{\sqrt{N_B}} \nonumber \\
&=& \sqrt{N_{\psi}} \frac{\epsilon^2 \sqrt{BR(\psi \rightarrow
 P\gamma)}BR(U\rightarrow l^+l^-)}{\sqrt{|F_{\psi P}(q^2)|^2 \times [\mbox{QED}(q^2)]  \delta
 q^2}}
 \label{eq:ratio}
\end{eqnarray}
Therefore, with 1 billion $\psi$ events, the reach for U-boson
searching can be $\epsilon \sim 10^{-2}$ - $10^{-3}$ in the $\psi
\rightarrow PU$ decays. In Figs.~\ref{eep} and \ref{mumup}, we show
the reach of the parameter $\epsilon$ by defining $S/\sqrt{B} = 5$
for different mass of $U$ boson.

\section{Summary}

In summary, the EM Dalitz decays of $\psi \rightarrow P l^+l^-$ are
studied in this paper. We demonstrate the differential decay rate as
both $q^2$-dependent rate and angular dependent rate explicitly. By
assuming simple pole approximation the decay rates for $\psi
\rightarrow P l^+l^-$ are estimated for the first time. The
estimated Dalitz decay rates could reach $\sim 10^{-7}$ for $\psi
\rightarrow \pi^0 l^+l^-$ and $\sim 10^{-5}$ for $\psi \rightarrow
\eta l^+l^-$ and $\eta^\prime l^+l^-$, respectively. They will be
accessible in the BESIII experiment with data sample of 1 billion
$\psi$ decay events. Especially, the $q^2$-dependent differential
decay rate can be measured by looking at the invariant mass of the
lepton pairs.

In the BESIII experiment, these measurements will be important for
us to understand the interaction of vector charmonium states with
photon, as well as to probe new physics beyond the standard model.
we have investigated the signatures of a hidden $U(1)_d$ sector at
the BESIII experiment in $\psi\rightarrow PU$ decays, and find that
the BESIII should have an intrinsic sensitivity to the kinetic
mixing parameter $\epsilon$ in the range of $10^{-2}$ - $10^{-4}$,
which depends on the mass values of the $U$ boson.

One of the authors (H.~B.~Li) would like to thank Jianping Ma,
J\'er\^ome Charles and S\'ebastien Descotes-Genon for helpful
discussions. This work is supported in part by the National Natural
Science Foundation of China under contract Nos. 11125525,
10575108, 10975077, 10735080, and by the Fundamental Research Funds for the
Central Universities No. 65030021.





\begin{thebibliography}{99}

\bibitem{landsberg1982} L.~G. Landsberg, Sov. Phys. Usp. {\bf 28}
(1985)435.
\bibitem{landsberg1985} L.~G. Landsberg,  Phys. Rep. {\bf 128} (1985) 301.
\bibitem{kroll1955} N.~M. Kroll, W. Wada, Phys. Rev. {\bf 98} (1955) 1355.
\bibitem{sov1992} N.~N. Achasov, A.~A. Kozhevnikov, Sov. J. Nucl. Phys. {\bf 55} (1992) 449.
\bibitem{iva2011} S.~A. Ivashyn, arXiv:1111.1291[hep-ph].
\bibitem{le2010} C. Terschlusen and S. Leupold, Phys. Lett. {\bf B691} (2010) 191.
\bibitem{faessler2000} A. Faessler, C. Fuchs, M.~I. Krivoruchenko, Phys. Rev. {\bf C61} (2000)
035206.
\bibitem{klingl}F. Klingl, N. Kaiser, W. Weise, Z. Phys. {\bf A356} (1996) 193.
\bibitem{kopp} G. Kopp, Phys. Rev. {\bf D10} (1974) 932.

\bibitem{pdg2010} K.~Nakamura, {\it et al.}, (Particle data group), J.
of Phys. {\bf G37} (2010) 1.
\bibitem{na60-2009} R. Arnaldi, {\it et al.}, NA60 Collaboration, Phys. Lett. {\bf B677} (2009) 260.
\bibitem{snd-phi} M. N. Achasov, {\it et al.}, SND Collaboration, Phys. Lett. {\bf B504} (2001) 275.
\bibitem{kloe-hadron11} C. D. Donato, {\it talk at hadron 2011},
arXiv:1109.3968 [hep-ex].

\bibitem{dark-1} C. Boehm, D. Hooper, J. Silk, M. Casse, J. Paul, Phys. Rev.
Lett. {\bf 92}, 101301 (2004).
\bibitem{dark-2} S.~N. Gninenko, N.~V. Krasnikov,
Phys. Lett. B {\bf 513}, 119(2001).
\bibitem{dark-3} P. Fayet, Phys. Rev. D {\bf 70},
023514 (2004); C. Bouchiat, P. Fayet, Phys. Lett. B {\bf 608}, 87
(2005).
 \bibitem{dark-4} A.~E. Dorokhov, M~.A. Ivanov, Phys. Rev. D {\bf 75}, 114007(2007);
 Y. Kahn, M. Schmitt, T.~M.~P. Tait, Phys. Rev. D {\bf 78}, 115002(2008).
 \bibitem{dark-5} S.~H. Zhu, Phys. Rev. D {\bf 75}, 115004 (2007); P.~F. Yin, S.~H. Zhu,
 Phys. Lett. B {\bf 679}, 362(2009).
 \bibitem{dark-6} N. Borodatchenkova, D. Choudhury, M. Drees, Phys.
Rev. Lett. {\bf 96}, 141802 (2006).
\bibitem{u-1} N. Arkani-Hamed, D.~P.
Finkbeiner, T.~R. Slatyer, N. Weiner, Phys. Rev. D {\bf79}, 015014
(2009); N. Arkani-Hamed, N. Weiner, JHEP {\bf 0812}, 104 (2008).
\bibitem{u-2} B. Holdom, Phys. Lett. B {\bf 178}, 65 (1986).
\bibitem{u-3} K.~R. Dienes, C.~F. Kolda, J. March-Russel, Nucl. Phys. B {\bf 492},
104(1997).

\bibitem{haibo-hadron11} H.~B.~Li, {\it talk at hadron2011}, arXiv:1108.5789 [hep-ex].

\bibitem{fayet1} P.~Fayet, Nucl. Phys. {\bf B 187} (1981) 184; Phys.
Lett. {\bf B} (1980)285.
\bibitem{fayet2} P.~Fayet, Phys. Rev. {\bf D 74} (2006) 054034.
\bibitem{fayet3} P.~Fayet, Phys. Rev. {\bf D 75} (2007) 115017.
\bibitem{li2010} H.~B.~Li and T.~Luo, Phys. lett. {\bf B 686}
(2010)249.

\bibitem{kloe} F. Archilli, {\it et al.}, KLOE-2 Collaboration, arXiv:1110.0411[hep-ex].
\bibitem{gellman61} M.~Gell-Mann and F.~Zachariasen, Phys. Rev.
{\bf 124} (1961) 953.
\bibitem{bauer78} R.~P.~Feynman, Photon-hadron interaction
(Benjamin, New York, 1972); T.~H.~Bauer {\it et al.},
Rev.~Mod.~Phys. {\bf 50} (1978) 261.

\bibitem{bepcii} J.~Z.~Bai {\it et al.} (BES Collaboration),  Nucl. Instrum. Meth. A {\bf
344}, 319 (1994); J.~Z.~Bai {\it et al.} (BES Collaboration),  Nucl.
Instrum. Meth. A {\bf 627}, 319 (2001).
\bibitem{wang} M. Reece, L.T. Wang, JHEP 0907,051 (2009).
\end{thebibliography}
\end{document}